\documentclass[runningheads]{llncs}

\usepackage{graphicx}
\usepackage{listings}
\usepackage{xcolor}

\colorlet{punct}{red!60!black}
\definecolor{background}{HTML}{EEEEEE}
\definecolor{delim}{RGB}{20,105,176}

\lstdefinelanguage{json}{
    basicstyle=\ttfamily\footnotesize,
    numbers=left,
    numberstyle=\scriptsize,
    stepnumber=1,
    numbersep=5pt,
    captionpos=b,
    showstringspaces=false,
    breaklines=true,
    frame=lines,
    backgroundcolor=\color{background},
    literate=
      {:}{{{\color{punct}{:}}}}{1}
      {,}{{{\color{punct}{,}}}}{1}
      {\{}{{{\color{delim}{\{}}}}{1}
      {\}}{{{\color{delim}{\}}}}}{1}
      {[}{{{\color{delim}{[}}}}{1}
      {]}{{{\color{delim}{]}}}}{1},
}

\begin{document}
\title{Implementation of the IIIF Presentation API 3.0 based on Software Support}
\subtitle{Use Case of an Incremental IIIF Deployment within a Citizen Science Project}
%
%
\author{Julien Antoine Raemy\inst{1}\orcidID{0000-0002-4711-5759} \and \\
Adrian Demleitner\inst{1}\orcidID{0000-0001-9918-7300}}
\authorrunning{J. A. Raemy and A. Demleitner}
%
\institute{Digital Humanities Lab, University of Basel, Spalenberg 65, Basel, Switzerland\\
\email{julien.raemy@unibas.ch}, \email{adrian.demleitner@unibas.ch} \\ 
\url{https://dhlab.philhist.unibas.ch/}}
\maketitle              

\begin{abstract}
As part of the Participatory Knowledge Practices in Analogue and Digital Image Archives (PIA) research project, we have been implementing Linked Open Usable Data (LOUD) standards including the International Image Interoperability Framework (IIIF) specifications to disseminate digital objects, their related metadata and streamline our processes. We have taken an incremental approach to IIIF deployment, first by installing the Simple Image Presentation Interface (SIPI), a IIIF Image API 3.0 server, followed by conceiving a workflow based on cookbook recipes created and vetted by the IIIF community for the generation of resources compatible with the IIIF Presentation API 3.0, one of the key components of our architecture. This workflow resulted in a monitoring exercise of this community-driven effort, principally to align the requirements of PIA and the IIIF Presentation API support of software clients.

\keywords{Citizen Science \and Cultural Heritage \and International Image Interoperability Framework \and Linked Open Usable Data \and Participatory Knowledge Practices in Analogue and Digital Image Archives \and Swiss Society for Folklore Studies.}
\end{abstract}
\begin{description}
    \item[Preprint] This is the preprint version of a conference paper that was accepted at EuroMed2022, the International Conference on Digital Heritage, that took place in Limassol, Cyprus between the 7th November and the 11th November 2022. The conference proceedings are due to be published by Springer Nature Publisher in the Lecture Notes in Computer Science (LNCS) series.
\end{description}

\section{Introduction}
Citizen Science projects and initiatives in the Humanities (sometimes referred to as Citizen Humanities \cite{heinisch_citizen_2021}) have become increasingly popular over the last fifteen years thanks to digital transformation \cite{ridge_2_2021}, whether via platforms for enriching or correcting existing metadata (crowdsourcing \cite{simpson_zooniverse_2014}, transcription) or in participatory forms such as calls for images or for input from a wider and more diverse public. From a technological point of view, the disruption will be rather in the curation of data and the support of participants as well as how to integrate or publish enriched or new material and their related metadata than in the development of new norms \cite{ridge_8_2021}. 

Within the Participatory Knowledge Practices in Analogue and Digital Image Archives (PIA) project, we intend to rely on standards that have already proven themselves in the digital heritage field \cite{raemy_suggested_2019}, in particular the International Image Interoperability Framework (IIIF), in order to break down silos between institutions and also to facilitate the exchange of objects and their integration for cultural heritage institutions or universities, while at the same time having digital resources that can be easily exploited beyond the lifetime of a research project, whether machine-readable or human-readable \cite{raemy_applying_2021}.

This requires us to focus on the minimal technical requirements of Citizen Science initiatives in order to perpetuate knowledge and also to target the level of compliance with the specifications. In our opinion, this should be done at the level of the client software, which often only partially supports the standards as there is limited value in promoting standards and disseminating compliant objects in Citizen Science if only a handful of users will know how to access or interpret such content. In other words, a bottom-up technological approach for the very initiatives that are supposed to be so, to prevent these data from becoming residual categories \cite{star_this_2010} that are difficult to interpret in the long-term \cite{vohland_science_2021}.

\section{Participatory Knowledge Practices in Analogue and Digital Image Archives (PIA)}
PIA\footnote{\url{https://about.participatory-archives.ch/}} is a four-year research project initiated in February 2021 by the Institute for Cultural Anthropology and European Ethnology and the Digital Humanities Lab of the University of Basel together with the Institute of Design Research of the Bern Academy of the Arts. It connects the world of archival data and analogue things in an interdisciplinary manner and aims to explore the phases of the analogue and digital archive from the perspectives of cultural anthropology, technology and design. Digital tools that support contextualising, linking and contrasting images will be developed. 

The project is dedicated to the collaboration of the scientific community and the wider public, facilitating the preservation and dissemination of knowledge, and encouraging users to engage collaboratively with their own history and contemporary practices. Thus, it is as much a matter of embedding the project in a citizen science context as it is of pursuing partnerships with various academic institutions and projects. PIA draws on three collections of the photographic archives\footnote{The three collections are SGV\_05 \textit{Atlas der Schweizerischen Volkskunde}, SGV\_10 \textit{Familie Kreis}, and SGV\_12 \textit{Ernst Brunner}. The SSFS Photo Archive website is available at: \url{https://archiv.sgv-sstp.ch/}.} of the Swiss Society for Folklore Studies (SSFS), comprising images on topics such as everyday life, tradition, identity, or labour. 

PIA has endowed itself with a common vision emphasising seven priorities or pillars: \textit{accessibility}, \textit{heterogeneity}, \textit{materiality}, \textit{interoperability}, \textit{affinites}, \textit{artificial intelligence} and \textit{bias management}. In order to make PIA part of the Open Science movement and also to meet the priorities outlined previously (notably \textit{accessibility}, \textit{interoperability} and \textit{affinities}), the technologies and software will wherever possible be open source. Furthermore, we want to build on standards that are in line with the Linked Open Usable Data (LOUD) principles\footnote{Similar to Tim Berners Lee's 5-star deployment scheme for Open Data, five design principles underpin LOUD: \url{https://linked.art/loud/}.} for the description (Linked Art), dissemination (IIIF) and annotation (Web Annotation Data Model) of the digital objects~\cite{raemy_ameliorer_2022}. 

\section{The International Image Interoperability Framework (IIIF)}
The International Image Interoperability Framework (IIIF) is a global community that has created and maintained shared application programming interfaces (APIs) to display and annotate digital representations of objects~\cite{snydman_international_2015}. It has been bringing together important actors from the academic and cultural heritage fields for over ten years and formed a consortium (IIIF-C) in 2015 to sustain the initiative and support its uptake~\cite{raemy_international_2017}.

\subsection{The Core IIIF APIs}
IIIF has defined six APIs\footnote{\url{https://iiif.io/api/}}. The two most important IIIF APIs, often referred to as the core APIs are the Image and Presentation APIs, both of which are in version 3 since June 2020. Hereafter both APIs are briefly outlined.

\subsubsection{The IIIF Image API} specifies a RESTful web service that returns an image in response to a standard HTTP(S) request \cite{appleby_iiif_2020-1}. It can be called in the following two ways:

\begin{enumerate}
    \item \textbf{Image Request}: Request of an image, derived from the underlying image content. The characteristics of the returned image can be manipulated through five parameters: \texttt{region}, \texttt{size}, \texttt{rotation}, \texttt{quality}, and \texttt{format}.
    \item \textbf{Image Information}: Request information about the image service, including characteristics, functionality available, and related services. The syntax of the response is serialised in JavaScript Object Notation for Linked Data (JSON-LD).
\end{enumerate}

Table~\ref{tab_imageapi} shows the IIIF Image API Request URI Syntax.

\begin{table}
\centering
\caption{IIIF Image API Syntax}\label{tab_imageapi}
\begin{tabular}{|l|l|}
\hline
Base URI          & \{scheme\}://\{server\}\{/prefix\}/\{identifier\}                  \\
Image Request     & \{\$BASE\}/\{region\}/\{size\}/\{rotation\}/\{quality\}.\{format\} \\
Image Information & \{\$BASE\}/info.json \\
\hline
\end{tabular}
\end{table}

\subsubsection{The IIIF Presentation API} is a JSON-LD based web service which provides the necessary information about the object or collection structure and layout. The purpose of the API is to display descriptive information that is intended for humans and does not aim to provide semantic metadata for search engines \cite{appleby_iiif_2020}.

\begin{figure}
\centering
\includegraphics[width=8cm]{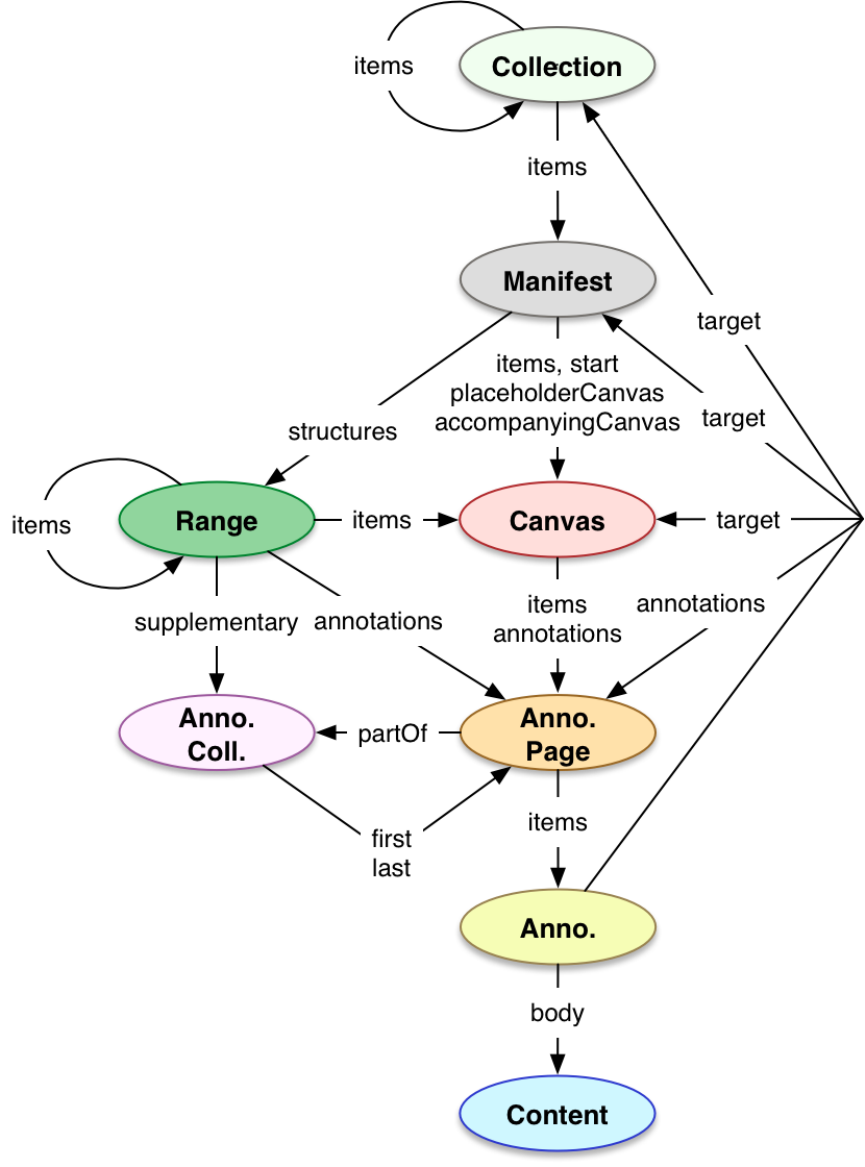}
\caption{IIIF Presentation API 3.0 Data Model \cite{appleby_iiif_2020}} \label{fig_prezidatamodel}
\end{figure}

The specification is loosely derived from the Shared Canvas Data Model \cite{sanderson_sharedcanvas_2011} which consists of the following four main types of resources: \texttt{Collection}, \texttt{Manifest}, \texttt{Canvas}, \texttt{Range} (see Fig.~\ref{fig_prezidatamodel} which highlights the relationships between all resource types).

 \subsection{IIIF-compliant Software}
In order to deploy IIIF, the system building blocks must understand the different IIIF APIs. In other words, to implement a basic IIIF solution, the images must be delivered in accordance with the Image API via a web service\footnote{Indeed, this implies that the resources to be delivered by tiles - or even on-the-fly - through a server are image-based. The IIIF Presentation API also works without the IIIF Image API, with static images, or when dealing with audio or video content.}, and then a series of scripts or micro-services must be leveraged to generate any IIIF Presentation API Resources (e.g. \texttt{Manifests}), bringing together the relevant metadata and data structure, which can be then displayed within a IIIF-compliant software client (or viewer) which supports both core APIs\footnote{There are also a number of viewers that only supports the IIIF Image API, like OpenSeadragon or Leaflet-IIIF. }, such as:

\begin{itemize}
\item \textbf{Mirador} - \url{https://projectmirador.org}
\item \textbf{Universal Viewer (UV)} - \url{http://universalviewer.io/}
\item \textbf{Annona} - \url{https://ncsu-libraries.github.io/annona/multistoryboard/}
\item \textbf{Clover} - \url{https://samvera-labs.github.io/clover-iiif/}
\end{itemize}

Deploying a IIIF architecture that can manage and save annotations, limit access to certain objects, inform the IIIF community of changes made within its own collection or be able to search for optical character recognition (OCR) text requires a more complex environment, notably by deploying several types of servers and different libraries. 
In this paper, we give an example of the IIIF workflow that we have been creating for the PIA research project in Sect. \ref{pia_iiif}, which could also prove relevant to similar initiatives.

\subsection{A Community-driven Effort}
IIIF is a community-driven initiative that over the years has become more structured and professional, e.g. with the creation of a consortium\footnote{\url{https://iiif.io/community/consortium/}}, a code of conduct and by employing three staff members.

Various (technical or community) working groups, (temporary or ongoing) committees meet regularly online to carry forward thematic discussions, to organise upcoming conferences or workshops, or to decide on the guidelines and overall strategy of IIIF. Some of these groups are open to all and others are more or less restricted to representatives of the consortium. This is for instance the case of the Technical Review Committee (TRC) where every full member of IIIF-C has a place but only a handful of people from the wider community can sit on a limited basis. This group plays a very significant role in evaluating the work of the editors (yet another key player in the IIIF community) in approving APIs and extensions but also cookbook recipes, which have become an increasingly important component for implementing the IIIF Presentation API 3.0 (more on this in the next subsection).

\begin{figure}
\includegraphics[width=\textwidth]{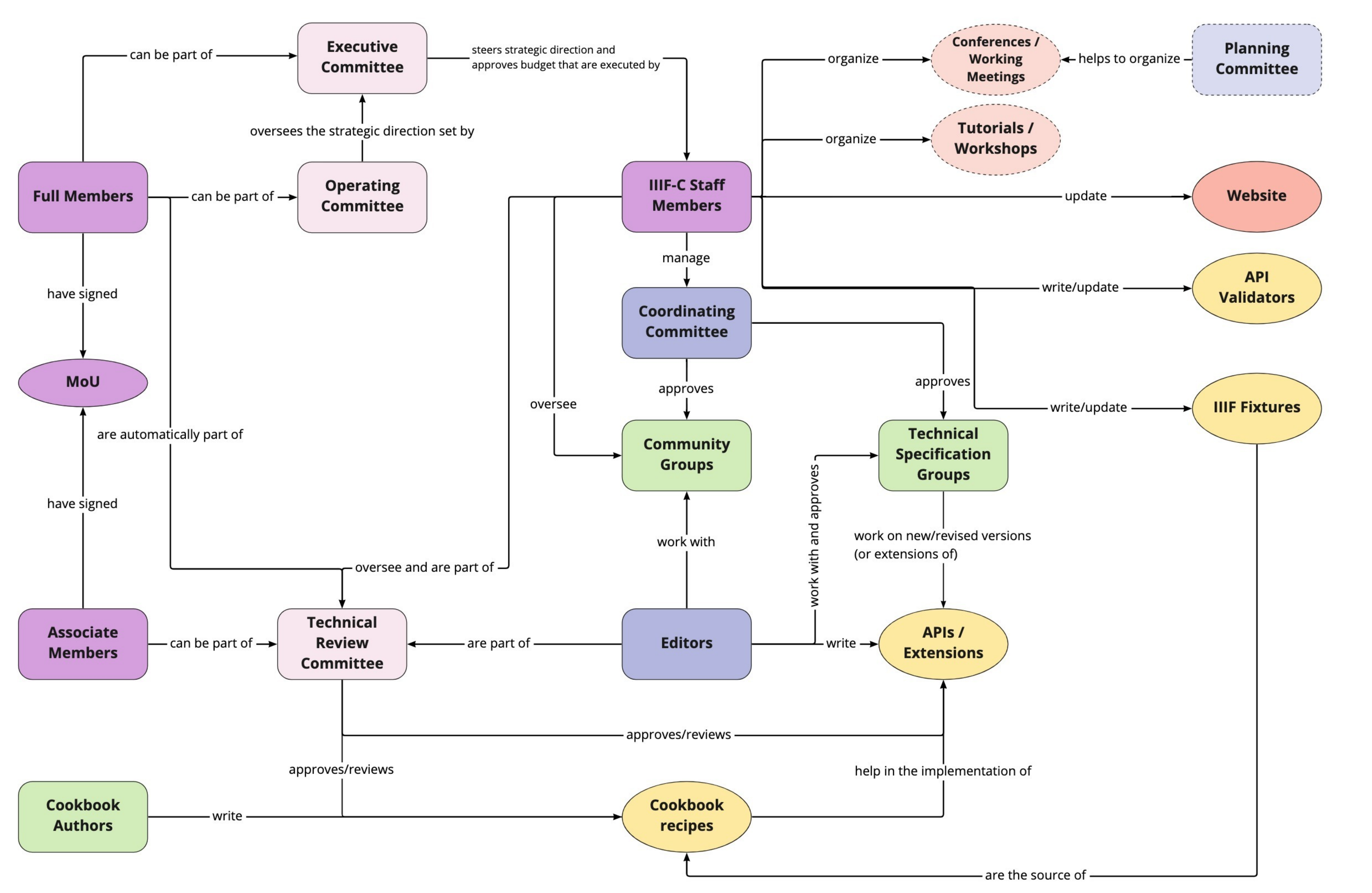}
\caption{Overview of the IIIF Community} \label{fig_iiif_community}
\end{figure}

Figure~\ref{fig_iiif_community} illustrates the intricate connections between the different actors of the IIIF community and their respective involvement.

\subsection{The IIIF Cookbook}
After the release of the IIIF Presentation API 3.0, the IIIF community has been keeping a Cookbook of recipes highlighting different patterns (by types of content, by properties, by topic, etc.) within IIIF resources (mostly Manifests) to give examples of implementation, to show the diversity of IIIF use cases or to encourage people publishing IIIF resources to follow these recipes which are vetted by the TRC. The Cookbook process also helps on highlighting and converging use cases not yet supported by implementers. 

Each recipe is structured as follows: \textit{use cases},\textit{ implementation notes}, \textit{restrictions}, \textit{examples}, followed by the JSON-LD representation of a given IIIF resource where the described pattern is highlighted (see Listing \ref{lst:la} retrieved from \url{https://iiif.io/api/cookbook/recipe/0053-seeAlso/}. If the recipe is supported by one or more viewers, appropriate links are provided as well.

\begin{lstlisting}[label={lst:la},caption=Snippet from the \textit{Linking to Structured Metadata} Cookbook recipe, language=JSON]
  "seeAlso": [
    {
      "id": "https://fixtures.iiif.io/other/UCLA/ezukushi_mods.xml",
      "type": "Dataset",
      "label": {
        "en": [
          "MODS metadata"
        ]
      },
      "format": "text/xml",
      "profile": "http://www.loc.gov/mods/v3"
    }
  ],
\end{lstlisting}
Alongside this effort, a viewer matrix\footnote{https://iiif.io/api/cookbook/recipe/matrix/} is maintained to indicate the support (\textit{yes}, \textit{partial}, \textit{no}) of the various software clients in relation to the highlighted pattern(s). It must be said that viewers may display a IIIF resource without supporting or rendering properly a given pattern and in many cases with no disclaimer, i.e., similar to what happens across different software, the user experience (UX) will vary significantly\footnote{To this end, we can mention the work of the IIIF Design Community Group which seeks to mitigate such outcomes and seeks to provide UX best practices within the IIIF ecosystem: \url{https://iiif.io/community/groups/design/}.}.

As of September 2022, 42 unique cookbook recipes have been approved by the TRC and three software clients (Mirador, the UV, and Annona) are listed. If Mirador is the viewer with the widest support, the following should be noted: 

\begin{itemize}
\item Only eight out of 42 recipes (19\%) have no support across these three viewers. This figure rises to nine if partial support is excluded. 
\item Each viewer has its strengths and related background: Mirador was conceived as a workspace for comparison and annotation purposes. Meanwhile UV is much more versatile in terms of handling audio-visual content as well as 3D\footnote{Currently out-of-scope but it is foreseen to have such integration in future IIIF APIs.}, whereas Annona is as much an annotation tool as a storytelling application and supports more recent recipes, especially in relation to geographical metadata.
\end{itemize}

\begin{figure}
\includegraphics[width=\textwidth]{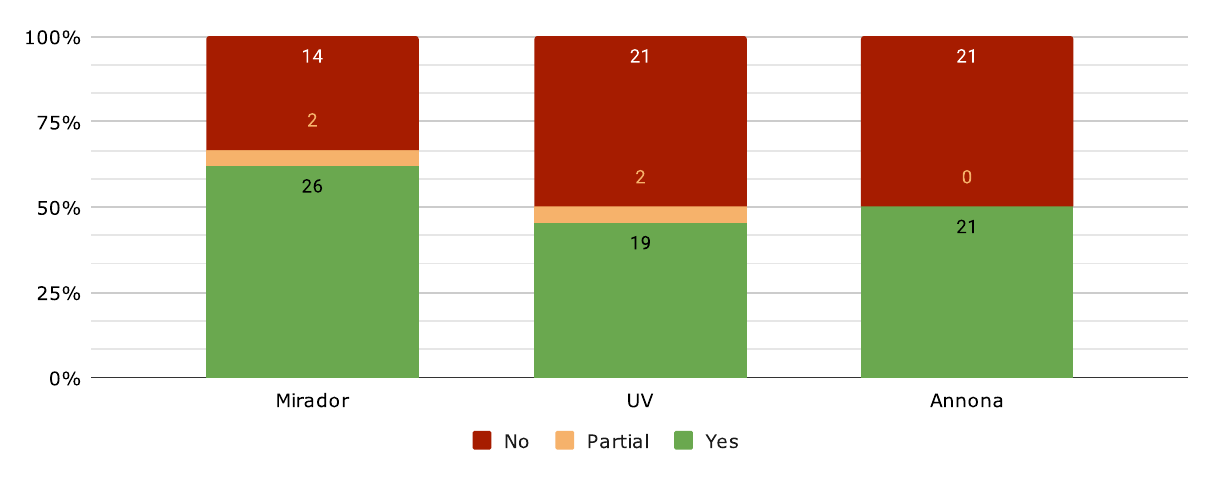}
\caption{Viewer support of the IIIF Cookbook recipes (Sept. 2022) \cite{raemy_back_2022}} \label{viewersupport}
\end{figure}
\section{Implementation of IIIF within PIA} \label{pia_iiif}
As mentioned earlier, PIA aims to leverage open systems and in particular standards in line with the LOUD design principles, such as IIIF.  First of all, the Simple Image Presentation Interface (SIPI), a C++ web server compliant with the IIIF Image API 3.0 was deployed \cite{rosenthaler_simple_2017}. SIPI has been developed and is maintained at the Digital Humanities Lab of the University of Basel, which greatly influenced our choice as we have a privileged relationship with the main developer. Secondly, we had to think about how to create and disseminate IIIF resources compatible with the IIIF Presentation API 3.0, as although there are general microservices or scripts available, it was more convenient to create something ad-hoc, again relying on an open framework.

\subsection{The PIA IIIF Workflow}
We have developed an application for IIIF Resources based on Laravel, an open-source PHP Framework\footnote{\url{https://codeberg.org/PIA/pia-iiif-manifest-host}} for generating IIIF resources (Manifests and Collections) as well as machine-learning Annotations, which are derived from vitrivr, a content-based multimedia retrieval system \cite{gasser_multimodal_2019}. To serve IIIF manifests, the application consumes metadata from our repositories. 

Our main database is managed through Omeka S, which offers its own JSON-LD API. The data that we gather via this API is combined with the machine-learning Object Detection which are hosted in a separate SQLite database. 

\begin{figure}
\includegraphics[width=\textwidth]{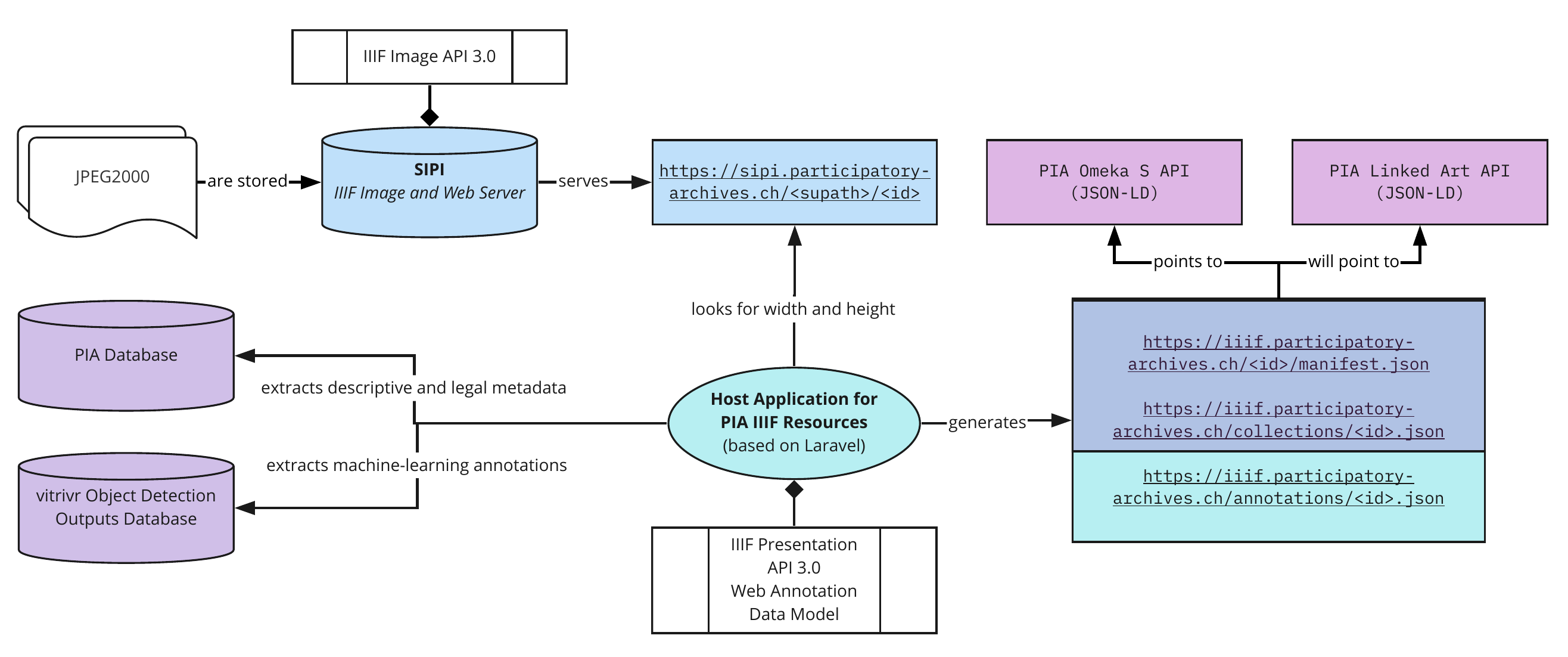}
\caption{PIA IIIF Resource Workflow} \label{fig_pia_iiif}
\end{figure}

Finally, we enrich our temporary metadata collection with image-specific information provided by SIPI, such as image-dimensions. The application then translates this data-construct into the IIIF specific format and serves the generated JSON file to the client (see Fig.~\ref{fig_pia_iiif}).

To build our IIIF resources, we have created boilerplates (or templates) that bring together many of the patterns presented in the cookbook recipes. Table \ref{cookbook_relevance} summarises which recipes we have incorporated into our \texttt{Manifest} boilerplates:

\begin{table}
\centering
\caption{Relevance of the IIIF Cookbook against our resources}
\label{cookbook_relevance}
\resizebox{\textwidth}{!}{%
\begin{tabular}{|l|l|}
\hline
\textbf{IIIF Cookbook recipe}                                                                                & \textbf{Relevance}                                                                                                       \\ \hline
\begin{tabular}[c]{@{}l@{}}Simplest Manifest - Single Image File \end{tabular} & \begin{tabular}[c]{@{}l@{}}The majority of the SSFS photographic archive\\  is made up of individual images\end{tabular} \\ \hline
\begin{tabular}[c]{@{}l@{}}Simple Manifest - Book  \end{tabular}                                            & Photo albums                                                                                                             \\ \hline
\begin{tabular}[c]{@{}l@{}}Support Deep Viewing \\ with Basic Use of a IIIF Image Service\end{tabular}       & Leveraging our SIPI instance capabilities                                                                                \\ \hline
Internationalization and Multi-language Values                                                               & Metadata in all official Swiss languages                                                                                 \\ \hline
Acknowledge Content Contributors (\texttt{providers})                                                                 & Giving credits to the SSFS and the participants                                                                          \\ \hline
Linking to Structured Metadata (\texttt{seeAlso})                                           & \begin{tabular}[c]{@{}l@{}}Pointing to semantic metadata \\ for aggregation or discovery purposes\end{tabular}           \\ \hline
Embedded or referenced Annotations                                                                           & \begin{tabular}[c]{@{}l@{}}Pointing to one AnnotationPage per user \\ (including Machine-Learning outputs)\end{tabular}  \\ \hline
\end{tabular}%
}
\end{table}

\subsection{Integration of IIIF into our user interface}
Out of the generic requirements for developing a Citizen Science infrastructure in terms of technology and standards \cite{raemy_applying_2021}, we noticed that those related to IIIF were still vague, especially with regard to its integration into our user interface. 

\begin{figure}
\includegraphics[width=\textwidth]{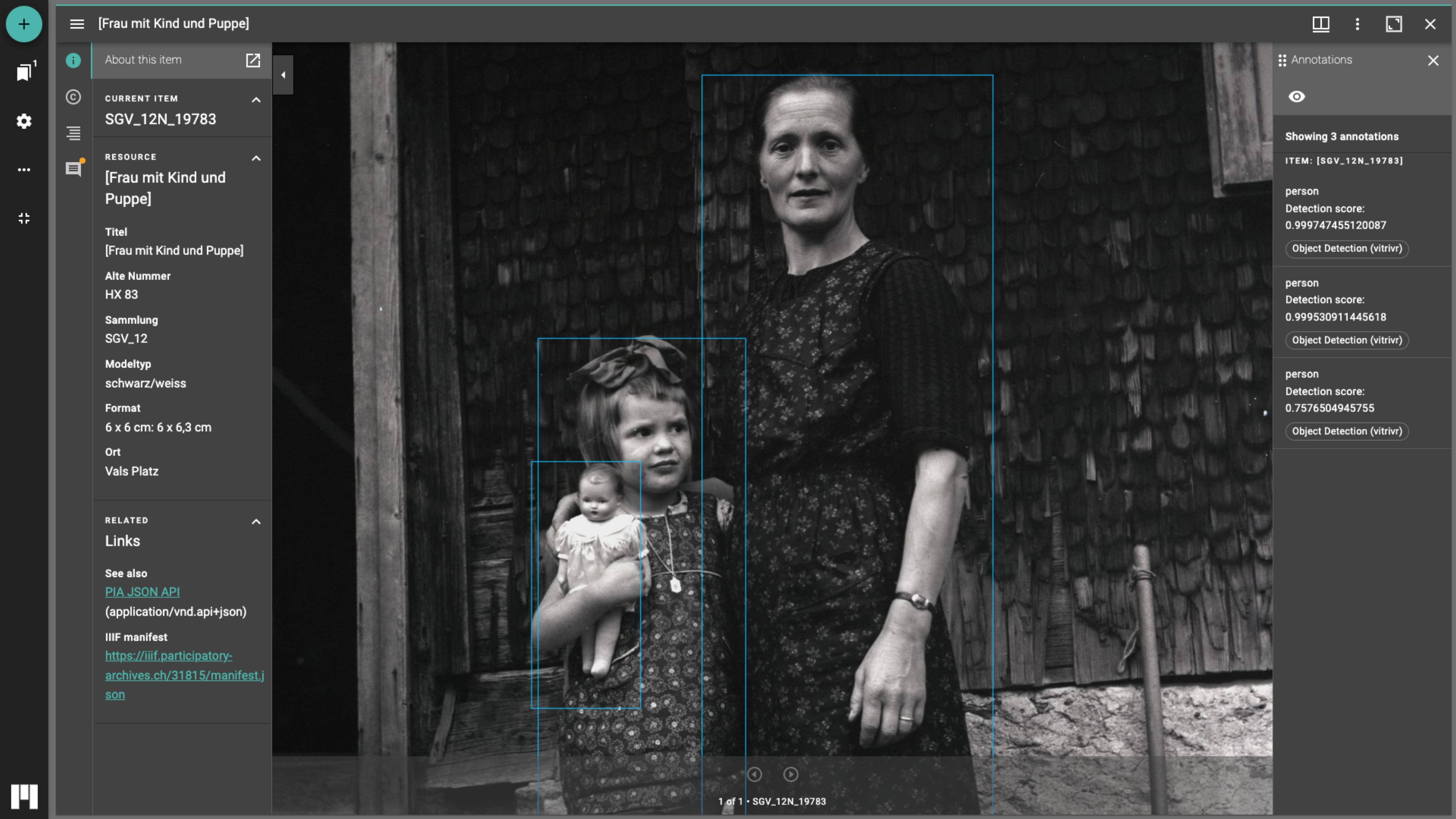}
\caption{PIA Mirador instance displaying a IIIF Manifest based on a digitised photograph of the Ernst Brunner Collection.} \label{fig_mirador}
\end{figure}

Thus we identified a number of key elements. For instance, when an image or a part of an image has to be retrieved, i.e. for the creation of thumbnails, for some search hits or even for the creation of memes, the IIIF Image API will be leveraged without the use of a viewer. As for the display of images alongside their metadata, Leaflet-IIIF was selected as we already have other flavors of Leaflet integrated in other applications. For object comparison, displaying annotations, as well as manipulating images, we decided to opt for Mirador (see Fig.~\ref{fig_mirador} which displays a IIIF Manifest within our Mirador instance). 

\section{Discussion and Future Work}
By positioning viewers as \textit{moderators}, and not mere \textit{intermediaries} to borrow a Latourian terminology \cite{latour_reassembling_2005}, within distributed systems, they have a significant acting role to play, as much for developers, scholars as for the general public. Their gain can also only be realised if a seamless incorporation is kept in place \cite{zundert_not_2018}. This may be wishful thinking in the long run, but it is to be hoped that the IIIF community, consisting mainly of technologists (admittedly usually from libraries or museums) and a few scholars, will continually consider the multifaceted challenges of user requirements in a bottom-up approach while continuing to maintain and develop easy to adopt specifications.

New cookbook recipes will be integrated into PIA, such as relevant patterns facilitating the display of geographical metadata within or pointing to maps. This monitoring work is indeed simplified as one of us is a member of the IIIF TRC, but a more automated setting without necessarily going through each recipe or the viewer matrix should in our opinion be sought. Solutions are being discussed within the IIIF community\footnote{\url{https://github.com/IIIF/cookbook-recipes/issues/100}} to report on software support, or at least to aggregate recipes more easily, for example by providing an Activity Streams of each recipe via the IIIF Change Discovery API \cite{appleby_iiif_2021}. Another noteworthy development is the work of Jie Song and his team who have incorporated cookbook recipes as well as Mirador and the UV into a command-line tool called etu-cli\footnote{\url{https://github.com/etu-wiki/etu-cli}}, which also facilitates the workflow but requires the owners to update their repository regularly in terms of IIIF resources and compatible viewers.

\section{Conclusion}
We have put forward an incremental integration of IIIF within the PIA research project based on viewer support (or on the expectation that unsupported patterns will soon be covered by a push from the wider IIIF community). We regard the IIIF APIs, and more broadly those that adhere to the LOUD design principles, as facilitators, whether it be for a sense of belonging within a community that is able able to work collaboratively on use cases as well as for its simplicity for displaying complex objects both for humans and machines.

The benefits of deploying such open and interoperable standards seem obvious to us, what still needs to be assessed is how to further improve the data stewardship and sustainability of such content and metadata in the context of Citizen Science.

\subsubsection{Acknowledgements.}
The PIA research project is financed by the Swiss National Science Foundation (SNSF) as part of their Sinergia programme (contract no. CRSII5\_193788): \url{https://data.snf.ch/grants/grant/193788}.

%
%

%
%
%


\end{document}